\documentclass{PoS}

\title{On the highest energy emission from millisecond pulsars}

\ShortTitle{Highest energy emission from MSPs}

\author{\speaker{P.~M.~Saz Parkinson}\\
        Santa Cruz Institute for Particle Physics, University of
        California, Santa Cruz, CA, USA\\
        Dept. of Physics \& Laboratory for Space Research, The University of Hong Kong, Hong Kong\\
        E-mail: \email{pablosp@hku.hk}}

\author{Andrea Belfiore\\
Istituto di Astrofisica Spaziale e Fisica Cosmica Milano, via E. Bassini 15, 20133, Milano, Italy}

\author{David Carreto Fidalgo\\
Grupo de Altas Energ\'{i}as (GAE), Universidad Complutense, 28040
Madrid, Spain}

\author{Colin J. Clark\\
Albert-Einstein-Institut, Max-Planck-Institut f\"{u}r
Gravitationsphysik, Hannover, Germany\\
Leibniz Universit\"{a}t Hannover, D-30167 Hannover, Germany\\
Jodrell Bank Centre for Astrophysics, School of Physics and Astronomy, The University of
Manchester, Manchester M13 9PL, UK}

\author{Matthew Kerr\\
Space Science Division, Naval Research Laboratory, Washington, DC
20375-5352, USA}

\author{Lars Nieder\\
Albert-Einstein-Institut, Max-Planck-Institut f\"{u}r
Gravitationsphysik, Hannover, Germany\\
Leibniz Universit\"{a}t Hannover, D-30167 Hannover, Germany}

\author{on behalf of the {\slshape Fermi} LAT Collaboration}

\abstract{{\it Fermi} has detected over 200 pulsars above 100 MeV. In a previous work, using 3 years of LAT data (1FHL catalog) we reported
that 28 of these pulsars show emission above 10 GeV; only three of
these, however, were millisecond pulsars (MSPs). The recently-released Third Catalog of Hard {\it Fermi}-LAT Sources (3FHL) contains over 1500 sources
showing emission above 10 GeV, 17 of which are associated with
gamma-ray MSPs. Using three times as much data as in our previous
study (1FHL), we report on a systematic analysis of these pulsars
to determine the highest energy (pulsed) emission from MSPs and discuss the best possible candidates
for follow-up observations with ground-based TeV instruments (H.E.S.S., MAGIC, VERITAS, and the upcoming CTA).}

\FullConference{7th Fermi Symposium 2017\\
		15-20 October 2017\\
		Garmisch-Partenkirchen, Germany}

\begin{document}

\section{Introduction}
Studies of the $\gamma$-ray sky above 10 GeV were limited in the past
by the relatively poor sensitivity of the instruments. The Energetic
Gamma Ray Experiment Telescope (EGRET) on the {\it Compton Gamma Ray Observatory (CGRO)} detected just over 1500 photons above 10 GeV in its 9-year
lifetime, 37 of which fell within 1$^\circ$ of one of the 5
EGRET-detected $\gamma$-ray pulsars known at the
time~\cite{Thompson05}. The {\it Fermi} Large Area Telescope (LAT),
launched in 2008, has already exceeded EGRET's lifetime and with its
larger field of view and better sensitivity, has collected about three
orders of magnitude more photons than EGRET did. Based on
silicon-strip detector technology, compared to the older generation
gas-based spark chambers, the LAT's improved sensitivity is
particularly significant at the highest energies. The First {\it Fermi}-LAT
Catalog of sources above 10 GeV (1FHL)~\cite{1FHL}, using the first three years
of LAT data above 10 GeV, contained 514 sources. Most of the 1FHL
sources (76\%) are associated with active galactic nuclei, but 27 are
associated with pulsars, while $\sim13\%$ of 1FHL sources are unassociated.
\section{Pulsars in 1FHL and 2FHL}
The 27 pulsars coincident with 1FHL sources covered every category,
including the 5 EGRET-detected pulsars, roughly equal numbers of young
{\it radio loud} and {\it radio quiet} pulsars, and 5 millisecond
pulsars (MSPs). A likelihood ratio analysis was carried out, comparing
the {\it low energy} ($>$100 MeV) light curve to the {\it high energy}
($>10$ GeV) events, in order to determine whether significant pulsations were
detected above 10 GeV and 25 GeV. A majority of the pulsars tested (20/25\footnote{PSR J1536--4948 and
  J2339--0533 were not part of the Second Fermi LAT Catalog of
  Gamma-ray Pulsars (2PC, ~\cite{2PC}), so these were left out of the
  analysis.}) were found to ``pulse'' above 10 GeV, with 12
pulsars showing significant pulsations above 25 GeV (including the MSP
J0614--3329). In addition to these, 14 pulsars from the 2nd Pulsar
Catalog which showed evidence for possible emission above 10 GeV were
studied. Of these, eight showed significant pulsations above 10 GeV,
including the MSPs J2017+0603 and J2302+4442. In short, \cite{1FHL}
showed that at least 28 gamma-ray pulsars show pulsations above 10
GeV, 5 of which were MSPs, with one MSP (J0614--3329) showing
pulsations above 25 GeV. 
The Second Catalog of Hard {\it Fermi}-LAT Sources (2FHL)~\cite{2FHL}
spanned 80 months and used the improved {\it Pass 8} data. However,
the low energy threshold this time around was set at 50 GeV, instead of 10
GeV used for 1FHL. This resulted in far fewer sources (360), and only
one of these was coincident with a pulsar (Vela). Indeed,
\cite{Leung14} confirmed that Vela emits pulsations above 50 GeV, something 
also confirmed now by H.E.S.S.~\cite{djannati17}.

\section{The 3FHL Catalog}
The Third Catalog of Hard {\it Fermi}-LAT Sources (3FHL)~\cite{3FHL} uses 7
years of {\it Pass 8} data above 10 GeV and contains
1556 sources. It's interesting to note that the number of $\gamma$-ray sources above
10 GeV now slightly exceeds the number of $>$10 GeV photons detected
by EGRET in its 9-year lifetime. Like the 1FHL before it, the 3FHL contains mainly (79\%) sources
associated with active galactic nuclei, with $\sim$13\% of its sources
being unassociated. 
\section{MSPs in 3FHL}
Compared to 1FHL, there are approximately three times the number of
pulsars associated with 3FHL sources, including 17 $\gamma$-ray
MSPs. Note that only 5 MSPs were previously coincident with 1FHL
sources. We have undertaken a similar analysis to that carried out in
1FHL, but this time focussed only on the MSPs associated with 3FHL
sources. 

The 17 MSPs coincident with 3FHL sources span a range of different
categories of MSPs: {\it isolated MSPs} such as J0340+4130 and
J0533+6759, the {\it Black Widow} pulsar J1311--3430, {\it
  Redback} MSPs, such as J2215+5135 and J2339--0533, and even the {\it
  transitional} MSP J1227--4853. We carried out a dedicated analysis
using 9 years of LAT data (three times the amount of data used in 1FHL). The
improved statistics enabled us to generate the {\it low energy} pulse
profile of each pulsar using events above 1 GeV. Because the pulse
profile of most pulsars is known to change with energy (with pulses
typically becoming narrower), this improves the sensitivity of the
pulsation search at higher energies. We performed a spectral analysis
of each of the pulsars, followed by an updated timing analysis of all
systems. We excluded J1227--4853 from our analysis for now, since its
pulsed emission is variable and requires a more careful dedicated
analysis. 

We find that all 16 MSPs that we tested showed significant
pulsations above 10 GeV. This includes PSRs J1536--4948 and
J2339--0533 which had been found to be coincident with 1FHL sources
but were not part of the previous analysis as these pulsars were not
part of 2PC~\cite{2PC}. Of these 16 MSPs, five show significant
pulsations above 25 GeV: J0218+4232, J0614--3329, J1231--1411,
J1311--3430, and J1536--4948~\cite{sazparkinson17}. 

{\bf PSR J0218+4232} was discovered over 25 years
ago~\cite{navarro95}. A tentative detection was claimed using EGRET
data~\cite{kuiper00} and it was among the first MSPs to be firmly
detected by the {\it Fermi} LAT~\cite{Abdo09b}. Its spindown luminosity of
2.4$\times10^{35}$ erg s$^{-1}$ makes it the second most energetic
{\it field} MSP\footnote{i.e. {\bf not} in a Globular Cluster.} (after J1939+2134), of the $\sim$100 gamma-ray MSPs detected so far by
{\it Fermi}. Its non-thermal X-ray flux, as well as its radio
luminosity ($L_{400}=S_{400} d^2_{kpc} \sim 400 $mJy kpc$^2$) are
among the highest of all MSPs~\cite{2PC}. {\bf J0614--3329}  and {\bf
  PSR J1231--1411} were two of the first three MSPs discovered in radio searches of {\it
  pulsar-like} unassociated $\gamma$-ray sources~\cite{ransom11} and
as such are among the brightest and most energetic MSPs detected by
the LAT ($\sim2\times10^{34}$erg s$^{-1}$). {\bf J0614--3329} was the
only MSP which was previously (in 1FHL) shown to have significant
pulsations above 25 GeV; in fact, despite being due to a single photon, the maximum energy above which the pulsed
emission was still found significant for this pulsar exceeded 60
GeV. Recently, \cite{Xing16} reported emission from this pulsar up to
60 GeV. The Black Widow pulsar {\bf J1311--3430} was the first binary MSP discovered in blind
searches of LAT data~\cite{Pletsch12}. Finally, {\bf PSR J1536--4948}
was discovered by the Giant Metrewave Radio Telescope (GMRT), in its
searches of {\it pulsar-like} LAT $\gamma$-ray
sources~\cite{Ray12}. 

It is indicative of the success of {\it
  Fermi}, that 4 out of the 5 $\gamma$-ray MSPs now known to be
emitting above 25 GeV were not even known prior to the launch of {\it
  Fermi}\footnote{J0614--3329, however, is coincident with an EGRET
  unidentified source.}.

\section{VHE Pulsars}
At energies above 50 GeV, the LAT unfortunately is simply too small to
collect sufficient statistics. Indeed, the total exposure on each of
the 16 MSPs described above, over the 9-year period of observation, ranges from
$\sim$1.0--1.7 m$^2$ yr, depending on the location in the
sky. Fortunately, ground-based $\gamma$-ray telescopes have made
tremendous advances in recent years. The MAGIC telescope led the way
with the first detection of the Crab pulsar above 25 GeV~\cite
{MAGIC_CrabI}, followed by the VERITAS detection of pulsed emission
above 100 GeV~\cite{VERITAS_Crab}. Most recently, MAGIC has detected
the Crab pulsar up to TeV energies~\cite{ansoldi16}. 
Added to the H.E.S.S. detection of the Vela pulsar up to 120
GeV~\cite{djannati17}, we now have two pulsars that emit at Very High
Energies (VHE, $>$100 GeV). A number of models have been developed to
explain this emission
(e.g. \cite{Aharonian12,Bednarek12,Lyutikov12,harding15}). Observationally,
it remains to be seen whether any other pulsars can be found that emit
at these energies. Current instruments like H.E.S.S., MAGIC, and
VERITAS continue to search for such VHE pulsars, and in a few years,
the much more sensitive Cherenkov Telescope Array (CTA) will join the
excitement, likely detecting a number of $\gamma$-ray pulsars,
starting at a few tens of GeV~\cite {Burtovoi17}.

\section{Summary}

The 1FHL Catalog~\cite{1FHL}, based on 3 years of LAT data above 10
GeV showed that 28 (13) pulsars showed significant pulsations above 10
(25) GeV, including 5 (1) MSPs emitting above 10 (25) GeV. The 3FHL
catalog includes 17 sources coincident with MSPs. We carried out an
analysis of 16 of these (excluding the {\it transitional} system
J1227--4853) and found that all 16 of them show significant pulsations
above 10 GeV, with 5 of them showing significant pulsations above 25
GeV. Detailed results are presented in \cite{sazparkinson17}.
Future obvservations with ground-based $\gamma$-ray telescopes like
H.E.S.S., MAGIC, and VERITAS, and in the near future CTA, will determine
whether any of these systems are VHE pulsars, emitting above 100 GeV,
like the Crab and Vela. The detection of new pulsars at these energies
will be crucial to test the many models that have been developed in recent
years to explain the VHE emission of the Crab and (to a lesser extent)
Vela pulsars.

\acknowledgments
The \textit{Fermi}-LAT Collaboration acknowledges support for LAT
development, operation and data analysis from NASA and DOE (United
States), CEA/Irfu and IN2P3/CNRS (France), ASI and INFN (Italy), MEXT,
KEK, and JAXA (Japan), and the K.A.~Wallenberg Foundation, the Swedish
Research Council and the National Space Board (Sweden). Science
analysis support in the operations phase from INAF (Italy) and CNES
(France) is also gratefully acknowledged. This work performed in part
under DOE Contract DE-AC02-76SF00515. This work was supported by the National Aeronautics and
Space Administration (NASA) through the Fermi Guest Investigator grant
NNX15AW43G. This research was partially carried out using the HKU Information Technology Services
research computing facilities that are supported in part by the Hong Kong
UGC Special Equipment Grant (SEG HKU09).

\def \apjl {ApJL}
\def \apj {ApJ}
\def \apjs {ApJS}
\def \jcap {JCAP}
\def \aap{A\&A}
\def \mnras{MNRAS}
\def \nat{Nature}
\def \apss{Astrophysics and Space Science}

\end{document}